\documentclass[runningheads]{llncs}
\usepackage[T1]{fontenc}
\usepackage{graphicx}
\usepackage[USenglish]{babel}
\usepackage{todonotes}
\usepackage{svg}
\usepackage{tabularx}
\usepackage{multirow}
\usepackage{pgfplots}
\pgfplotsset{compat=1.18}
\usepackage{pgf-pie}
\usepackage{subcaption}
\usepackage{enumitem}
\usepackage[framemethod=tikz]{mdframed}
\definecolor{link}{RGB}{0, 123, 255}
\usepackage[hidelinks]{hyperref}

\usepackage[nameinlink,capitalize]{cleveref}

\begin{document}
\title{Addressing Predicate Redundancy in Research Knowledge Graphs: Duplicate Detection, Resolution, and Prevention}
\titlerunning{Duplicate Detection, Resolution, and Prevention}
%
%
\author{Lena John\inst{1}\orcidID{0009-0007-2097-9761}\and
Sushant Aggarwal\inst{2}\and
Sören Auer\inst{1}\orcidID{0000-0002-0698-2864}\and
Oliver Karras\inst{1}\orcidID{0000-0001-5336-6899}
}
\authorrunning{L. John et al.}
\institute{
TIB - Leibniz Information Centre for Science and Technology, Hannover, Germany \and
Leibniz University Hannover, Germany
\\
\email{\{lena.john, soeren.auer, oliver.karras\}@tib.eu}\\
\texttt{sushant.aggarwal@stud.uni-hannover.de}
}
%

%
\maketitle              
\begin{abstract}
Research Knowledge Graphs (RKGs) enable the structured representation of scientific knowledge, but their weakly enforced schemas make them prone to inconsistencies, particularly in how predicates are defined and used. Duplicate predicates, i.e., distinct identifiers expressing the same or highly similar relationships, introduce semantic redundancy, hinder reuse, and reduce RKG quality. While prior work has addressed duplicate detection for downstream tasks such as query answering or schema alignment, predicate redundancy as a data quality challenge, remains underexplored, particularly in terms of resolution, prevention, and semi-automated curator support.

In this paper, we propose a framework for managing duplicate predicates in RKGs that covers detection, resolution, and prevention. The framework combines automated similarity-based methods with human validation and is designed for integration into the lifecycle of evolving, crowdsourced RKGs. We implement the framework in the context of the Open Research Knowledge Graph (ORKG) by extending its existing curation dashboard SciKGDash with embedding-based clustering, interactive inspection, and resolution actions such as merging and deleting.

We evaluate the framework on the ORKG, where clustering reveals that up to 30\% of predicates are potentially redundant. The analysis also shows recurring modeling patterns that lead to predicate redundancy, user-induced duplication, inconsistent identifier usage, and a lack of standardization in predicate naming and usage.

Our findings demonstrate that duplicate predicates arise from user behavior and interface design. Addressing this, requires combining automated methods with human-centered curation and preventive mechanisms. This work positions predicate redundancy as a central data quality challenge and provides a foundation for more systematic and proactive RKG curation.

\end{abstract}
\section{Introduction}
Knowledge Graphs (KGs) model entities and their relationships in a semantically enriched graph structure, enabling both human and machine interpretability~\cite{Hogan.2022}. Research Knowledge Graphs (RKGs) have emerged as a key technology for representing scientific knowledge across various domains~\cite{Stocker.2022}.

Their weakly enforced schemas and flexible nature support the integration of heterogeneous and evolving knowledge, but also introduce data quality challenges such as ambiguities, missing values, and duplicates~\cite{Hogan.2022,Huaman.2022,Huaman.2020}. Under the open world assumption (OWA)~\cite{Fürber.2011}, (R)KGs are inherently incomplete~\cite{Xue.2023}. Crowdsourced RKGs amplify these challenges by promoting decentralized and community-driven knowledge creation~\cite{Karras.2021}, which can lead to inconsistent modeling practices and evolving schemas~\cite{Setyawan.2020}. Ensuring semantic consistency therefore requires continuous curation beyond initial data integration~\cite{Huaman.2022}.

A central yet often overlooked source of inconsistency lies in predicates. While entities can often be aligned to external references, predicates define how these entities are semantically related and thus shape the meaning of (R)KGs. Divergent modeling practices frequently result in \textit{duplicate predicates}, i.e., distinct identifiers expressing the same or highly similar relationships~\cite{John_SciKGDash.2025,Li.2024}. These duplicates propagate semantic redundancy across all triples using them~\cite{Chen.2023}, reducing RKG quality and utility~\cite{Kalo.2020}. Despite advances in related areas such as ontology alignment~\cite{Babaei.2025,Zhao.2022} and schema mapping~\cite{Jeon.2025}, predicate redundancy remains largely underexplored as a data quality challenge.

This work addresses this challenge by proposing a framework for the detection, resolution, and prevention of duplicate predicates. 
The framework is informed by existing literature and practical curation experience in the Open Research Knowledge Graph (\href{https://orkg.org/}{\textcolor{link}{ORKG}})~\cite{Auer.2025}, a crowdsourced RKG for structured and machine-actionable scientific knowledge.
We implemented the framework within \href{https://scikgdash.orkg.org}{\textcolor{link}{SciKGDash}}~\cite{John_SciKGDash.2025}, the existing curation dashboard of the ORKG. We tested the detection and resolution actions of the framework in a sandbox instance of the ORKG that mirrors the live data, enabling safe experimentation and rollback while preserving real-world characteristics, and subsequently used the live system to derive practical insights into predicate redundancy and its curation.

Our findings indicate that a substantial portion of predicates (up to $30\%$) are potentially redundant and stem from recurring modeling patterns, including repeated creation of predicates with identical labels and descriptions but different identifiers by the same users, confusing identifiers with labels, and the absence of canonical forms for frequently used predicates.

The contribution of this work is twofold: (1) a structured framework for managing predicate redundancy in RKGs, and (2) a practical demonstration of its applicability in a real-world setting.\vspace{0.1cm}

The remainder of this paper is structured as follows: \cref{ch:Background} introduces the background knowledge. \cref{ch:RelatedWork} presents related work. \cref{ch:Framework} presents the framework. \cref{ch:Implementation} outlines the implementation. \cref{ch:Evaluation} presents the preliminary evaluation. \cref{ch:Discussion} discusses the findings. \cref{ch:Conclusion} concludes the paper, together with future work.

\section{Background: What is a duplicate predicate?}\label{ch:Background}
With duplicate predicates we refer to distinct predicates with different identifiers that encode the same or highly similar semantic relationship. We distinguish between syntactic similarity (labels, URIs), structural similarity (graph patterns), semantic similarity (equivalence in meaning and use), and ontological similarity (explicit formal relations). 
In this context, predicate uniqueness refers to the degree to which an RKG avoids multiple representations of the same underlying relation~\cite{Fürber.2011}.
When RKGs are populated, data (i.e. symbols) often transforms into information (i.e. connected data), and finally into knowledge (i.e. interpreted meaning)~\cite{Rowley.2007}. Duplicate predicates arise when information is not consistently abstracted into knowledge. Two predicates may differ at the data layer, but result into the same knowledge representation.

With regard to RKG quality, duplicate predicates fall under succinctness and more specifically conciseness, including both \textit{extensional conciseness} (redundant instances) and \textit{intensional representational conciseness} (multiple schema elements serving the same purpose)~\cite{Batini.2016,Hogan.2022}. Low conciseness ultimately leads to unnecessary proliferation of predicates without adding expressive power, thereby reducing overall structural clarity.

This issue is closely tied to a modeling trade-off between expressivity and abstraction. RKGs with fine-grained predicate sets increase expressivity and modeling flexibility, but introduce graph fragmentation and reduce predicate overlap, which negatively affects querying and inference~\cite{Chen.2023}. Conversely, more unified predicate sets improve overlap and reasoning capabilities, at the cost of semantic nuance.
This tension is further complicated by the OWA, under which missing information cannot be interpreted as false or complete. However, duplicate predicate detection requires a notion of completeness. Therefore, we adopt a closed world assumption (CWA) for this task, restricting the analysis to the observed predicate set and treating it as complete within the modeled scope~\cite{Fürber.2011}.

\section{Related Work}\label{ch:RelatedWork}
This section is based on an \href{https://doi.org/10.48366/R1637915}{\textcolor{link}{ORKG comparison}}~\cite{Comparison_John.2026} of 8 papers on approaches to duplicate detection, resolution, and prevention in KGs and Linked Data. ORKG comparisons provide a structured, tabular overview of contributions, enabling them to be compared along relevant properties.
The comparison shows that prior work primarily focuses on duplicate detection (7 out of 8) to support downstream tasks such as query expansion, schema alignment, and KG completion. For instance, Abedjan et al.~\cite{Abedjan.2013} perform schema analysis such as range-based filtering, while Kalo et al.~\cite{Kalo.2020,Kalo.2019} and Niazmand et al.~\cite{Niazmand.2024} apply rule mining and data-driven methods to identify semantically related predicates for query answering. Huaman et al.~\cite{Huaman.2020} are providing an overview for further duplicate detection tools.
In particular, ontology alignment approaches, e.g., within the \href{https://oaei.ontologymatching.org/}{\textcolor{link}{OAEI initiative}}, address schema similarity across ontologies and include methods for predicate matching based on lexical, structural, and embedding-based techniques~\cite{Babaei.2025,Chen.2023,Sousa.2023}. 
However, these approaches typically assume multiple ontologies and focus more strongly on entity alignment than on predicate redundancies within a single, evolving graph. Predicate matching remains challenging due to high lexical variability and heterogeneous representations~\cite{Sousa.2023}.

Although these methods identify semantically related predicates, they primarily treat similarity as a means to improve consumption tasks rather than as a data quality challenge, with few exceptions, such as Salem et al.~\cite{Salem.2022}. Approaches that explicitly consider duplicate resolution (2 out of 8) and prevention (2 out of 8) remain limited~\cite{Huaman.2020}. Notable examples include clustering-based predicate recommendation, based on paper title and abstract, by Oghli et al.~\cite{Arab.2022} in the ORKG and the GDup framework by Manghi et al.~\cite{Manghi.2020}, which supports detection, grouping, and merging with curator feedback. However, semi-automated approaches involving users or curators remain underexplored.

Overall, existing work rarely treats duplicate predicates as a first-class problem and lacks integrated support for detection, resolution, and prevention within the lifecycle of evolving, crowdsourced (R)KGs. In contrast, we address predicate redundancy as a data quality challenge targeting RKG curation and creation, and propose a reusable, human-centered framework covering all three stages.


\section{Framework}\label{ch:Framework}
To address predicate redundancy in especially crowdsourced RKGs, we propose an iterative framework consisting of Duplicate Detection (DD), Duplicate Resolution (DR), and Duplicate Prevention (DP). Based on our experience as contributors and curators of the ORKG, as well as our prior work on \href{https://scikgdash.orkg.org}{\textcolor{link}{SciKGDash}}~\cite{John_SciKGDash.2025} with the ORKG Curation \& Community Building team we encountered numerous duplicate predicates, which made it difficult to consistently reuse existing terminology. From a curation perspective, we identified it as a systemic issue: Users tend to create new predicates rather than reuse ones from the already noisy predicate set, further increasing redundancy. This motivated both the cleanup of existing predicates and the prevention of future duplicates, drawing on prior work on duplicate detection and resolution, especially by Manghi et al.~\cite{Manghi.2020}.

\begin{figure}[h]
    \vspace{-0.5cm}
    \centering
    \includegraphics[width=0.92\linewidth]{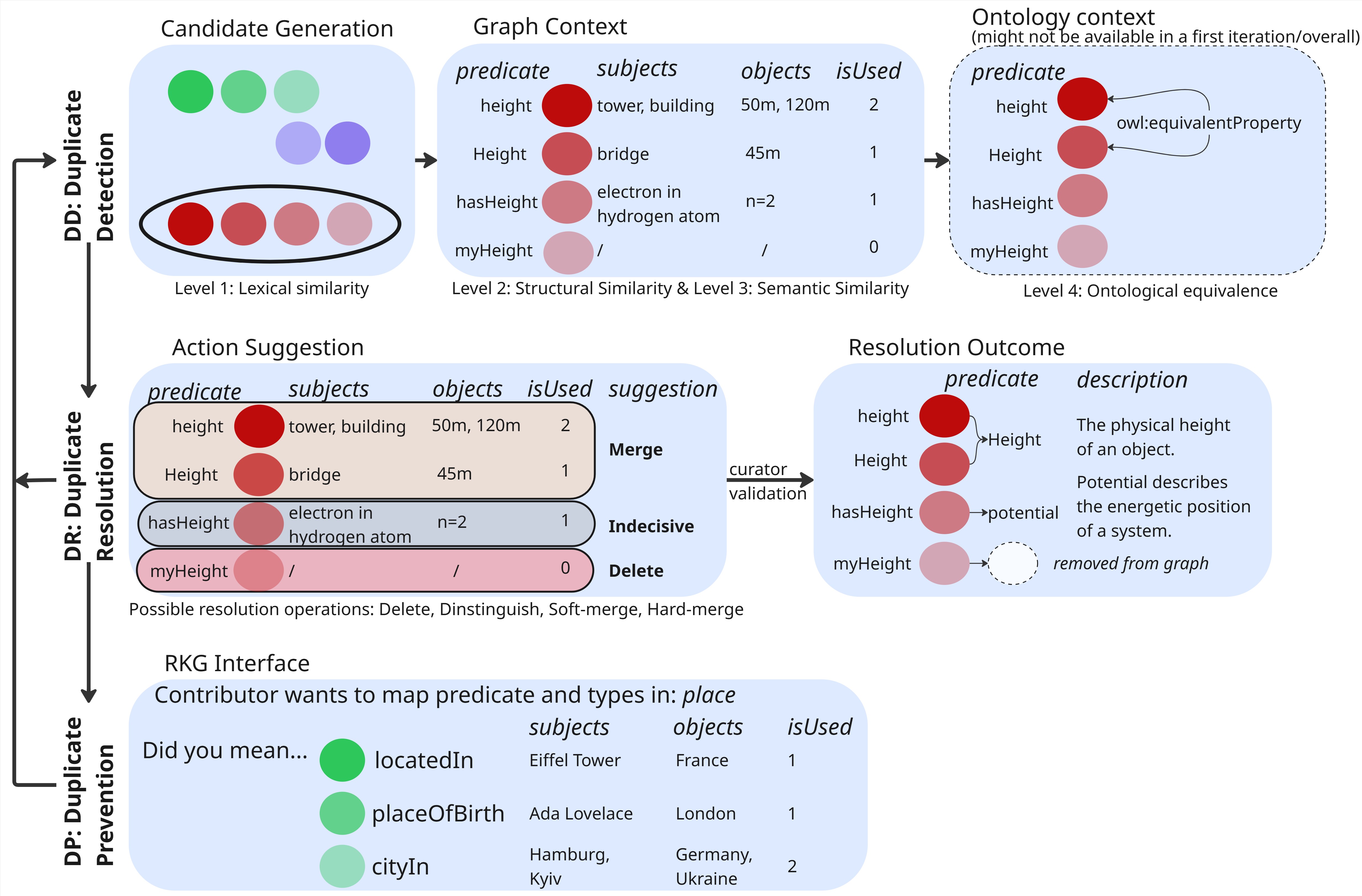}
    \caption{Framework for Duplicate Predicate Detection, Resolution, and Prevention. Detection uses stepwise similarity levels, resolution is guided by extensional usage and human validation, prevention is integrated into RKG import pipelines.}
    \label{fig:framework}
    \vspace{-1.0cm}
\end{figure}

The framework, as illustrated in \autoref{fig:framework}, follows an iterative process suited to continuously evolving RKGs with frequent data imports and contributions. DD and DR are inherently sequential, as duplicates must first be identified before they can be resolved, and are applied iteratively in practice. DD provides stepwise guidance for identifying duplicate predicates across four levels (cf. \autoref{ch:DD}), progressively incorporating richer contextual information. Ontological similarity may be absent in early iterations and can emerge through resolution decisions. DR suggests context-dependent resolution actions (cf. \autoref{ch:DR}), while final decisions remain under human control.
DP operates at a broader level and can be initiated at any time, but is most effective after DD and DR insights are available. These insights, such as root causes of duplication, usage patterns, and modeling behavior, inform targeted interventions in RKG design, user guidance, and data import processes. 
Resolving duplicates improves consistency, reducing the likelihood of future duplicate creation and prevention effort over time.
The framework combines automated support where possible with human oversight where necessary. It involves multiple roles~\cite{John_SciKGDash.2025,Li.2024}, that may overlap. Individuals may act as contributors, curators, and builders depending on the task at hand and level of automization. Contributors create content and potentially duplicates, curators analyze predicate usage and perform graph cleaning, builders refine interface design and modeling guidance. 

\subsection{Duplicate Detection (DD)}\label{ch:DD}
Duplicate detection can be formulated as a filtering problem with a precision–re\-call trade-off, supported by human curator validation. We define four levels ranging from lexical similarity to ontological equivalence (see \autoref{tab:duplicate_levels}), where each level provides increasing confidence in predicate equivalence. 
Candidate generation starts with Level~1 (lexical similarity), which is computationally efficient and maximizes recall by identifying predicates with similar labels or URIs. 
This step produces clusters of potential duplicates, but is prone to ambiguity and false positives due to polysemy. 
To improve precision, Levels~2 and~3 incorporate structural and semantic context, including subject/object types, graph usage patterns, and provenance. This filters out predicates that are similar at the surface level but differ in meaning, shifting from intensional similarity to extensional, usage-based similarity.
Since this stage often requires contextual interpretation, human validation is used to resolve ambiguous cases.
Level~4 leverages ontological information to identify explicit equivalences. However, such information is often unavailable a priori and may instead emerge as a result of resolution decisions in DR. Based on these results, refined candidate groups are passed to DR for final curator validation.

\begin{table}[hbt]
    \vspace{-0.5cm}
    \centering
    \caption{Levels of evidence for duplicate predicate detection. Higher levels indicate stronger justification for equivalence, lower levels may detect false positives.}
    {\footnotesize
    \begin{tabularx}{\textwidth}{p{0.24\textwidth}|X|X}
        Level \& Evidence& Likely Duplicates & False Duplicates \\\hline\hline
        
        \textbf{1. Lexical:}\par label/URI similarity
        & \texttt{method}/\texttt{Method}\par (naming variant) 
        & \texttt{location}/\texttt{location}\par (place vs. URL) \\\hline
        
        \textbf{2. Structural:}\par same graph usage
        & \texttt{Product hasPrice value.}\par \texttt{Product cost value.}\par (identical types) 
        & \texttt{Website head Title.}\par\texttt{Person headOf Company.}\par (different domains and ranges)\\\hline
        
        \textbf{3. Semantic:}\par contextual meaning
        & \texttt{Event period xsd:integer.}\par \texttt{Event length xsd:integer.} (used interchangeably)
        & \texttt{Sensor tempC xsd:decimal.} \texttt{Sensor tempF xsd:decimal.}\par (different units)\\\hline
        
        \textbf{4. Ontological:}\par formal equivalence
        & \texttt{foaf:name owl:equivalentProperty schema:name.}
        & \texttt{capitalOf rdfs:subPropertyOf locatedIn.} \\\hline
        
    \end{tabularx} 
    }
    \label{tab:duplicate_levels}
    \vspace{-1.0cm}
\end{table}

\subsection{Duplicate Resolution (DR)}\label{ch:DR}
The system proposes resolution actions for detected duplicate predicates. Curators select candidate predicates and apply one of four actions shown in~\autoref{tab:resolution_methods}: Delete, distinguish, soft-merge, or hard-merge. As these actions modify the RKG, they are restricted to authorized curators.
The \textit{delete} action removes unused predicates from the RKG. The \textit{distinguish} action refines labels or metadata to disambiguate semantically different predicates. For schemas, W3C standards do not provide explicit mechanisms to declare predicate inequality. OWL 2 \href{https://www.w3.org/TR/owl2-new-features/#F12:_Punning}{\textcolor{link}{punning}} allows predicates to be related via \texttt{owl:differentFrom}, but this is limited to individual instances and does not support reasoning across predicates.
The merge action is applied either as soft-merge or hard-merge. \textit{Soft-merge} introduces ontological equivalence (e.g., \texttt{owl:equivalentProperty} or \texttt{owl:sameAs} using punning) while preserving existing statements when confidence for hard-merge is limited. \textit{Hard-merge} replaces all occurrences of duplicate predicates with a canonical root predicate and removes redundant predicates from the RKG.

\subsection{Duplicate Prevention (DP)}\label{ch:DP}
DP begins already during the DD and DR phases by considering patterns. 
Because predicates form the semantic layer of an RKG, their extensional usage provides key insights into intended meaning. By analyzing creation and usage patterns, supported by workflow backtracking, usage analytics, and curation dashboards~\cite{John_SciKGDash.2025}, we can identify common causes of duplicate predicates.
Preventing duplicates therefore requires addressing these causes within the RKG interface and ingestion workflow, reducing the need for post-hoc DD and DR. 

Based on our prior~\cite{John_SciKGDash.2025} and present experiences, as well as general insights from literature~\cite{Huaman.2020}, typical sources of duplicate predicates include:

\begin{itemize}[leftmargin=*]
    \item \textbf{Limited visibility:} Existing predicates are hard to discover or browse.
    \item \textbf{Weak documentation:} Missing or unclear labels and usage guidance.
    \item \textbf{Low reuse incentives:} Creation is easier than reusing existing predicates. 
    \item \textbf{Insufficient safeguards:} Creation is not checked against similar predicates.
    \item \textbf{Import noise:} External data sources introduce duplicates during ingestion.
\end{itemize}
To mitigate these issues, RKG systems should provide real-time predicate suggestions, similarity-based warnings, usage analytics, and guided mapping during data import, especially during semantic enrichment phases~\cite{John_Scimantify.2025}. 

\begin{table}[!h]
    \vspace{-0.5cm}
    \centering
    \footnotesize
    \caption{Overview of resolution actions and their modifications to the RKG.}
    \begin{tabularx}{\textwidth}{c|p{0.19\textwidth}|p{0.32\textwidth}|X}
        \multirow{2}{*}{Action} & Example\par Predicate(s) & \multirow{2}{*}{Resolution action} & \multirow{2}{*}{Outcome} \\\hline\hline
        
        Delete & \texttt{tempLabel}\par (unused) & Remove predicate & Predicate is removed from the RKG (only if unused). \\\hline
        
        Distinguish & \texttt{degree} \par(temperature) vs. \texttt{degree}\par(qualification) & Refine labels or add metadata & Predicates remain, but are semantically disambiguated.\\\hline
        
        Soft-merge & \texttt{period}/\texttt{length}\par & Add \texttt{period}\par \texttt{owl:equivalentProperty}\par \texttt{length.} & Establishes ontological equivalence; candidates for later hard‑merge. \\\hline
        
        Hard-merge & \texttt{method}/\texttt{Method} & Replace with root predicate\par \texttt{Method $\rightarrow$ method, Method removed} & All instances mapped to one predicate; duplicates removed.\\\hline
        
    \end{tabularx}
    \label{tab:resolution_methods}
    \vspace{-1.0cm}
\end{table}

\section{Implementation}\label{ch:Implementation}
We implemented the proposed framework in the environment of the crowdsourced \href{https://orkg.org/}{\textcolor{link}{ORKG}} to demonstrate its practical applicability. To support DD and DR, we extend the curation dashboard \href{https://scikgdash.orkg.org}{\textcolor{link}{SciKGDash}}~\cite{John_SciKGDash.2025}, whose \href{https://gitlab.com/TIBHannover/orkg/scikgdash/-/tree/main?ref_type=heads}{\textcolor{link}{source code}} is publicly available. 
A dedicated \href{https://scikgdash.orkg.org/predicates}{\textcolor{link}{predicate curation interface}} integrates the DD and DR workflows.
DP is currently under development and planned for integration into the ORKG frontend and backend.

\href{https://scikgdash.orkg.org}{\textcolor{link}{SciKGDash}} consists of a Python backend, a PostgreSQL database, and a Next.js frontend. The DD pipeline retrieves predicates from ORKG, concatenates their labels and available descriptions into a single text sequence, embeds them using Sentence Transformers (\href{https://huggingface.co/sentence-transformers/all-MiniLM-L6-v2}{\textcolor{link}{all-MiniLM-L6-v2}}), and stores the results in the database.
Predicate similarity is computed via cosine similarity and a $k$-nearest neighbor graph ($k=50$, threshold $0.65$) is constructed to capture semantically related predicates.
These parameters were chosen based on exploratory experiments to balance connectivity among highly similar predicates with meaningful clustering.
Clustering is performed using \href{https://python-louvain.readthedocs.io/en/latest/}{\textcolor{link}{Louvain community detection}}, which ensures full assignment of predicates to clusters and scales efficiently to large graphs~\cite{Blondel.2008}, in contrast to density-based methods that may leave points unassigned.
To improve coverage of lexical duplicates beyond the $k$-neighborhood ($k=50$), predicates with identical normalized labels are explicitly linked prior to clustering.
An approximate global coherence score based on cluster sizes relative to the total number of predicates is displayed to motivate incremental curation progress.
For each cluster, lightweight heuristic rules provide low-cost initial curation suggestions. Predicates with zero usage and outdated creation timestamps are flagged for deletion, while identical labels indicate merge candidates. Remaining clusters are marked for manual review, enabling an initial screening before applying more expensive context-based similarity analysis.

\begin{figure}[h]
    \vspace{-0.5cm}
    \centering
    \fbox{\includegraphics[width=0.97\linewidth]{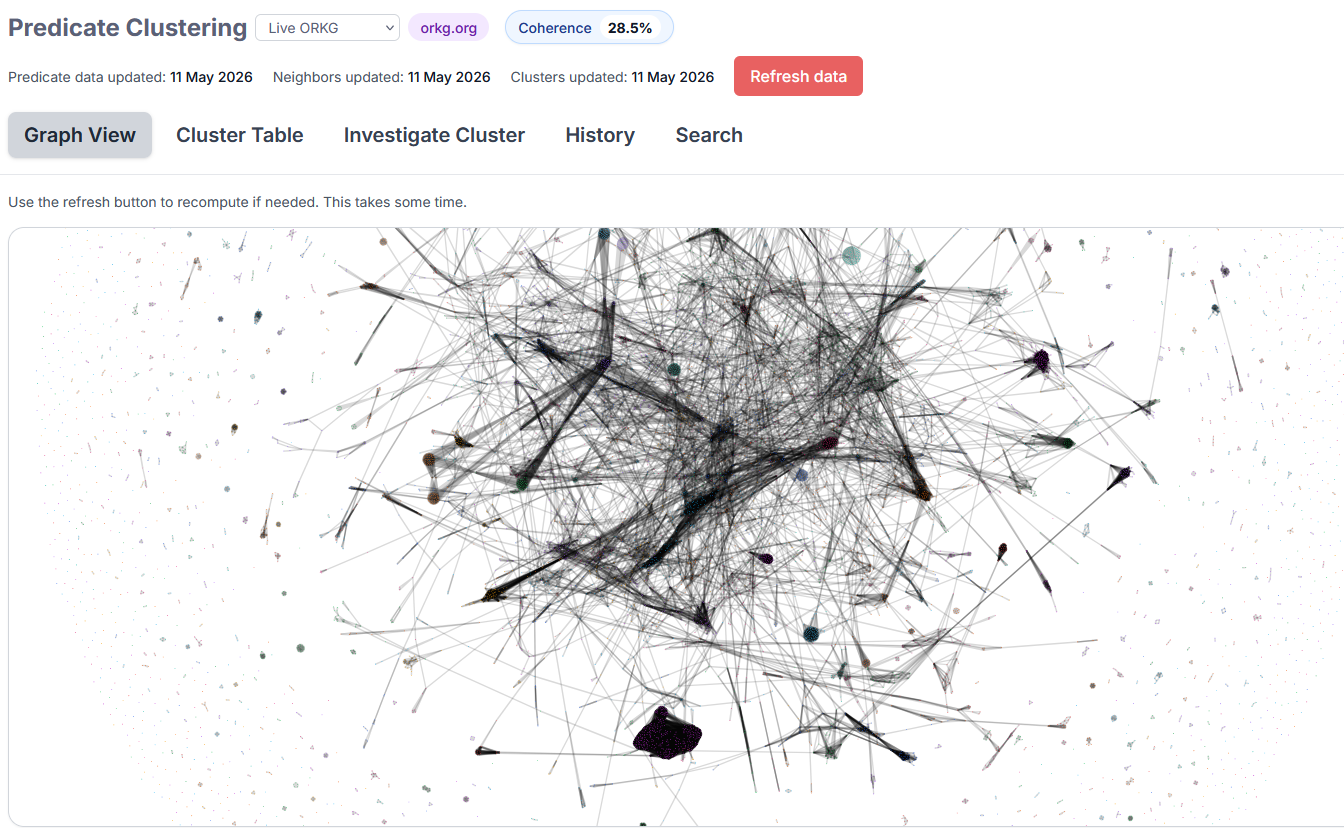}}
    \caption{Initial graph view of the $k$-nearest neighbor ($k=50$) similarity graph (cosine similarity of $0.65$) and its coherence score. Nodes represent predicates, edges indicate high semantic similarity, and clusters group potential duplicates.}
    \label{fig:clustering_graph_view}
    \vspace{-0.6cm}
\end{figure}

\begin{figure}[h]
    \centering
    \fbox{\includegraphics[width=0.98\linewidth]{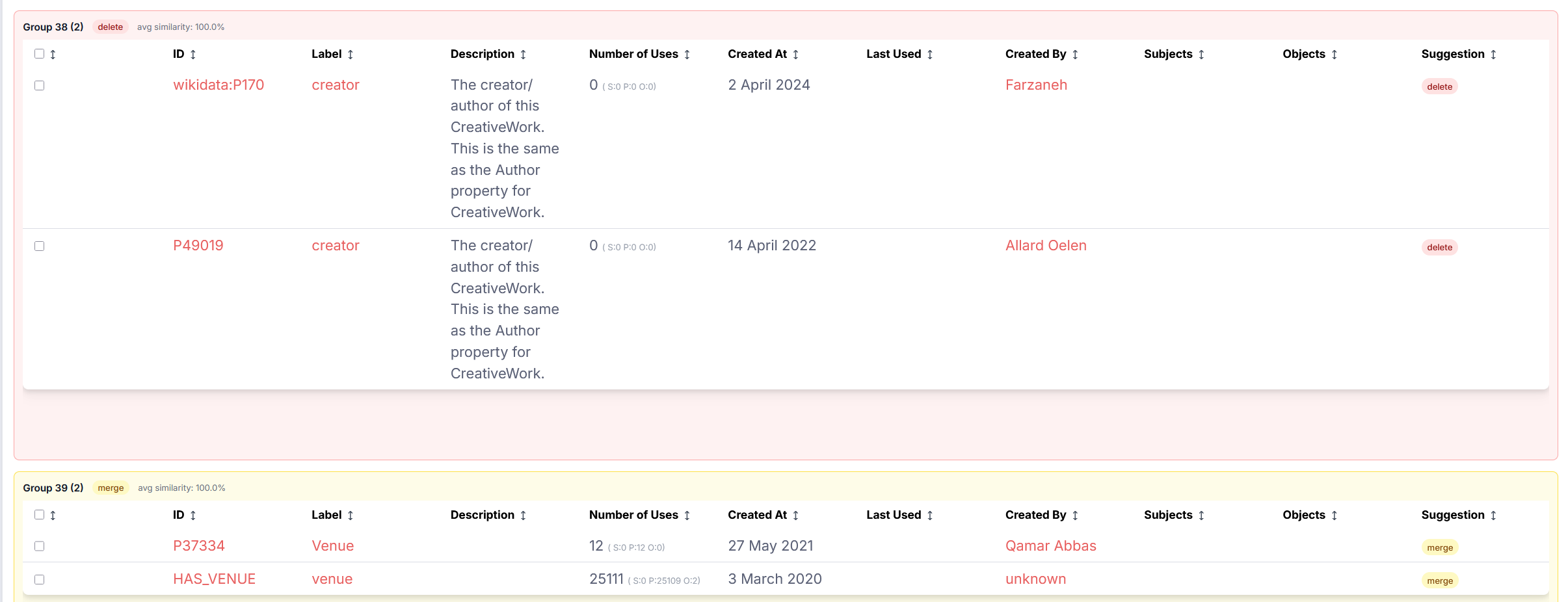}}
    \caption{Advanced suggestions with finer-grained groups based on contextual embeddings and higher similarity threshold ($0.8$) for a merge and a delete group.}
    \label{fig:groups_inside_clusters}
    \vspace{-0.5cm}
\end{figure}

Clusters are visualized as graph and list views. The graph view (see \autoref{fig:clustering_graph_view}) uses the \href{https://github.com/vasturiano/react-force-graph}{\textcolor{link}{react-force-graph library}}, which creates a two-dimensional network layout that produces dense visual structures with closely positioned nodes.
Curators can inspect cluster-based predicates and metadata, including its impact on the global coherence score, labels, descriptions, usage statistics, and linked subjects and objects. A key consideration specific to ORKG is, that predicates can occur in any position within a triple, which is reflected in the breakdown provided in the number of uses column (see \autoref{fig:groups_inside_clusters}).

An advanced suggestion mode refines a selected cluster by re-embedding predicates with the now available, extended contextual information of subjects and objects. Pairwise cosine similarity is then computed using a higher threshold ($0.8$), focusing on high-confidence semantic similarity, whereas the lower threshold ($0.65$) used in graph construction preserves broader connectivity. The resulting similarity graph in one cluster is partitioned into connected components via depth-first search, which are interpreted as fine-grained suggestion groups and labeled using heuristics based on usage, label similarity, and shared authorship, yielding candidate decisions such as merge, delete, or reviews (see \autoref{fig:groups_inside_clusters}).

The DR workflow allows authenticated curators to select predicates and apply actions on the production or sandbox ORKG instance. 
For merging, a root predicate is selected as the canonical target.
As ORKG does not employ a formal ontology and already uses the \texttt{sameAs} predicate, it is reused for soft-merges to declare equivalence.
Since \texttt{sameAs} does not enforce structural properties such as symmetry or transitivity~\cite{Baas.2021}, it is stored as a statement from the source predicate to the root.
Hard-merges additionally replace all occurrences of merged predicates in statements with the root predicate and remove the unused predicates.
Level 4 of DD is implemented prototypically. Ontological connections via \texttt{sameAs} are not yet visualized in the UI, but are nonetheless considered during hard-merges to prevent invalid self-references (e.g., \texttt{A sameAs A}).
Distinguish actions are currently supported via manual label and description editing in ORKG, while enhanced UI and (LLM-supported) metadata editing are planned.

Due to pipeline latency, there is a temporary discrepancy between backend and UI. Curator actions are recorded in a history tab and, until manual pipeline re-execution, modified predicates are filtered out to avoid inconsistent states. Automatically re-running the pipeline after each change was deemed inefficient.

\begin{figure}[h]
    \centering
    \begin{subfigure}[h]{0.45\textwidth}
        \centering
        \fbox{\includegraphics[width=1.2\linewidth,trim=0 2.28cm 0 2.28cm,
    clip]{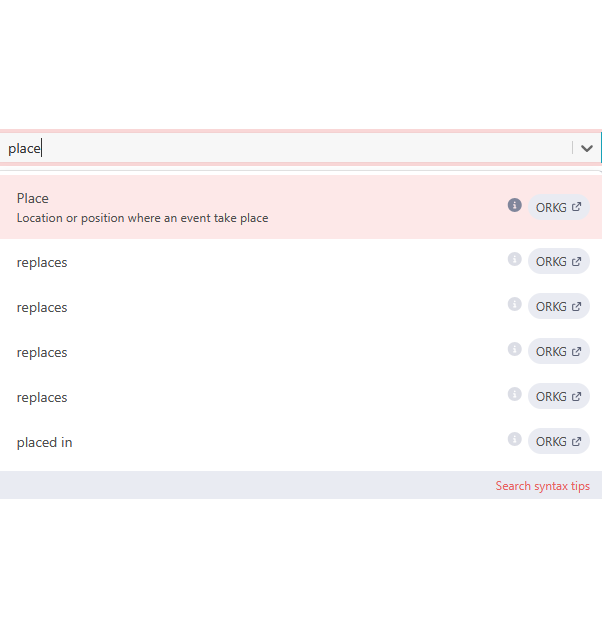}}
        \caption{Current ORKG search for predicates returning string match results.}
        \label{subfig:current_orkg_search}
    \end{subfigure}
    \hfill
    \begin{subfigure}[h]{0.4\textwidth}
        \centering
        \fbox{\includegraphics[width=0.9\linewidth]{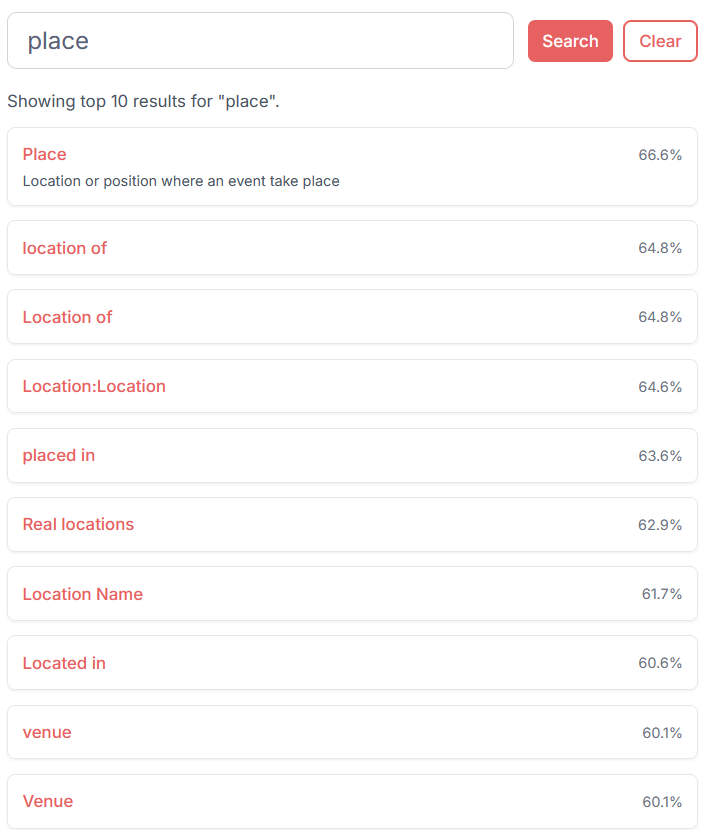}}
        \caption{SciKGDash prototype supporting similarity search results.}
        \label{subfig:similarity_search}
    \end{subfigure}
    \caption{Concept of embedding-based similarity search for duplicate prevention.}
    \label{fig:similarity_search_orkg}
    \vspace{-0.5cm}
\end{figure}

For DP, we aim to enhance the autocomplete for predicate selection, currently limited to string matching (see \autoref{subfig:current_orkg_search}), directly into the ORKG interface. We have \href{https://gitlab.com/TIBHannover/orkg/orkg-backend/-/tree/similarity-search-predicates?ref_type=heads}{\textcolor{link}{started extending}} the ORKG backend with \href{https://neo4j.com/docs/cypher-manual/current/indexes/semantic-indexes/vector-indexes/}{\textcolor{link}{Neo4j's vector index}} and plan to integrate \href{https://docs.spring.io/spring-ai/reference/api/embeddings/onnx.html}{\textcolor{link}{Sentence Transformers for Spring}}, enabling similarity-based predicate suggestions.
As ORKG is a mature system, these changes require coordination with the team due to the system’s complexity.
To explore the approach, a simplified version (see \autoref{subfig:similarity_search}) is implemented in the curation dashboard in the \href{https://scikgdash.orkg.org/predicates}{\textcolor{link}{search tab}}, where existing predicate embeddings are used.

\section{Evaluation}\label{ch:Evaluation}
\textbf{Analysis.} We conduct a feasibility study to evaluate the current state of predicate data in the ORKG and assess the potential of our framework for DD and DR.
The study outlines the technical capabilities of our implementation and focuses on identifying common modeling practices that lead to duplicate predicates, demonstrating how our approach can improve RKG quality. 
Since the framework operates on live graph data, a careful technical assessment is required before exposing it to general curators. 
Establishing a solid technical foundation is therefore a prerequisite for broader adoption by the ORKG curators.
To support the evaluation, we developed a \href{https://gitlab.com/TIBHannover/orkg/scikgdash/-/blob/develop/scikgdash-backend/scripts/impact_analysis.py}{\textcolor{link}{script}} that analyzes clustering results stored in the database. It computes cluster statistics, including size distributions and estimates of predicates eligible for merging or deletion. The corresponding raw results are \href{https://doi.org/10.5281/zenodo.19948623}{\textcolor{link}{online available}}~\cite{Zenodo_data.2026}.
We assume that an ideal similarity graph consists mostly of isolated nodes representing semantically distinct predicates, with only a small number of dense clusters indicating potential duplicates.\vspace{0.2cm}

Out of $13,457$ predicates, $3,837$ clusters were detected, including $2,769$ singleton clusters. A total of $2,801$ predicates appear as nodes without neighbors in the similarity graph but were later assigned to clusters based on lexical label similarity. This indicates that lexical signals can partially compensate for embedding limitations, particularly when descriptions distort the embedding space.
\autoref{tab:action_suggestion_distribution} shows the distribution of cluster-level recommendations derived from the advanced suggestions. Each cluster is treated as a single group and classified into \textit{merge}, \textit{delete}, \textit{review}, or \textit{keep} based on heuristic rules. Singleton clusters correspond to the \textit{keep} category, when they contain actively used predicates.
However, not all suggested actions are directly executable. The cleanup plan refines these results by applying additional constraints, such as excluding unused predicates and requiring at least two used predicates per merge candidate group. This reduces the set of actionable clusters.

\begin{table}[h]
    \vspace{-0.5cm}
    \caption{Distribution of cluster-level recommendations and affected predicates.}
    \centering
    \begin{tabular}{l|r|r|r|r}
    Action & {\shortstack{Clusters\\(Total)}} & {\shortstack{Clusters\\(Relative)}} & {\shortstack{Predicates\\(Total)}} & {\shortstack{Predicates\\(Relative)}} \\
    \hline\hline
    Merge  & 852  & $\approx$22.2\% & 3,185 & $\approx$23.7\% \\
    Delete & 338 & $\approx$8.8\% & 426 & $\approx$3.2\% \\
    Review & 162  & $\approx$4.2\% & 7,361 & $\approx$54.7\% \\
    Keep   & 2,485 & $\approx$64.8\% & 2,485 & $\approx$18.4\% \\
    \hline
    \
    {Total} & 3,837 & $\approx$100\% & 13,457 & $\approx$100\% \\
    \hline
    \end{tabular}
    \label{tab:action_suggestion_distribution}
\vspace{-0.5cm}
\end{table}

\begin{figure}[ht]
    \vspace{-0.5cm}
    \centering
    \begin{tikzpicture}
    \begin{axis}[
        ybar,
        bar width=12pt,
        grid=both,
        width=0.8\textwidth,
        height=0.3\textwidth,
        ylabel={Number of clusters},
        xlabel={Cluster size (binned)},
        symbolic x coords={1,2-5,6-10,11-50,51-100,100-477},
        xtick=data,
        ymin=0,
        ymax=3500,
        nodes near coords,
        nodes near coords align={vertical},
        enlarge x limits=0.15
    ]
    
    \addplot coordinates {
        (1,2769)
        (2-5,866)      
        (6-10,100)     
        (11-50,84)
        (51-100,8)
        (100-477,10)
    };
    
    \end{axis}
  \end{tikzpicture}
    \caption{Cluster size distribution grouped into bins. The distribution is highly skewed, with many singleton clusters and few large clusters.}
    \label{fig:cluster_distribution}
    \vspace{-0.5cm}
\end{figure}
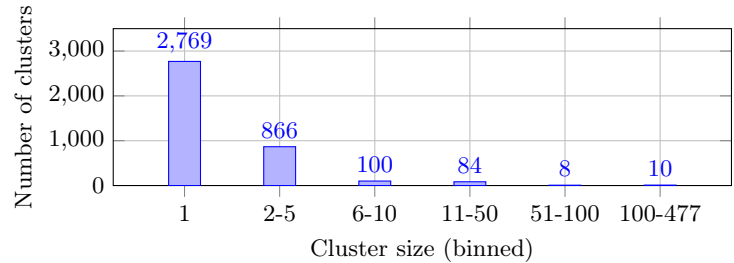

Overall, $1,989$ predicates ($14.78\%$) are unused and marked for deletion, also noting $54$ completely unused clusters. Furthermore, $2,044$ predicates ($15.19\%$) across $786$ clusters are identified as redundant merge candidates, meaning they can be replaced by a single representative predicate (already substracted) per cluster. In total, $4,033$ ($30\%$) predicates are considered potentially removable. However, these candidates require human validation, as computed similarity does not always imply true duplication.
Additional cluster-based signals provide further insight: $593$ clusters contain predicates created by the same user, and $79$ clusters contain identical labels. These patterns indicate recurring modeling inconsistencies that are further discussed below.
The cluster size distribution is highly skewed, with many singleton clusters and a small number of large clusters (up to size $477$), as shown in \autoref{fig:cluster_distribution}.

\noindent\textbf{Modeling patterns.} To complement the quantitative results, we manually inspected clusters for the live instance of ORKG using \href{https://scikgdash.orkg.org/predicates}{\textcolor{link}{SciKGDash}} and identified recurring modeling patterns that lead to duplicate or inconsistent predicates.\vspace{0.2cm}

\begin{figure}[h]
    \centering
    \begin{subfigure}[h]{0.46\textwidth}
        \centering
        \fbox{\includegraphics[height=4.28cm,keepaspectratio]{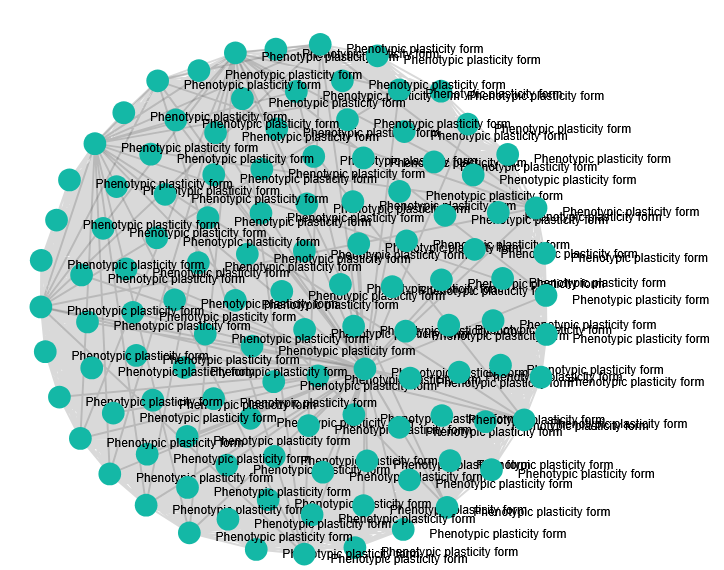}}
        \caption{Phenotypic plasticity form (size 115)}
        \label{fig:phenotypic_cluster}
    \end{subfigure}
    \hspace{0.3cm}
    \begin{subfigure}[h]{0.45\textwidth}
        \centering
        \fbox{\includegraphics[height=4.28cm,keepaspectratio]{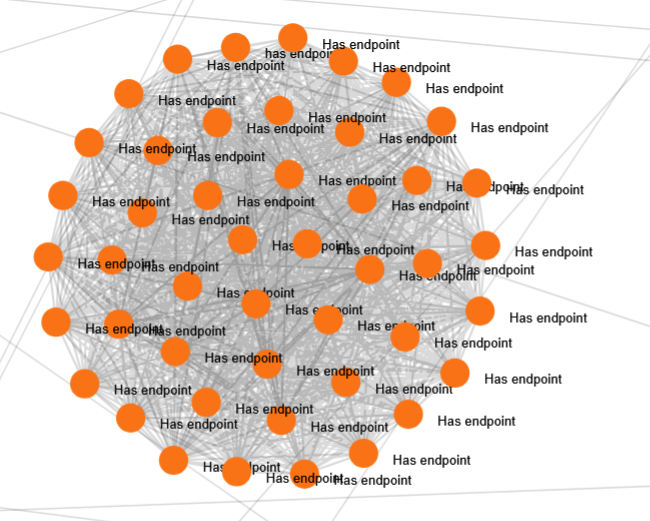}}
        \caption{Has endpoint (size 46)}
        \label{fig:has_endpoint_cluster}
    \end{subfigure} 
    \caption{Clusters arising from repeated predicate creation by the same user.}
    \label{fig:clusters_highly_connected}
    \vspace{-0.5cm}
\end{figure}

\noindent\textit{Repeated creation of identical predicates by a single user.} We observe cases (see \autoref{fig:clusters_highly_connected}) where the same user creates multiple (near-)identical predicates, often within short time spans. These form isolated, tightly connected clusters and represent clear duplication. Our framework successfully identifies these components and enables consolidation into a single canonical predicate.

\begin{figure}[h]
\vspace{-0.5cm}
\centering
\begin{minipage}{0.43\textwidth}
    \centering
    \fbox{\includegraphics[width=\linewidth]{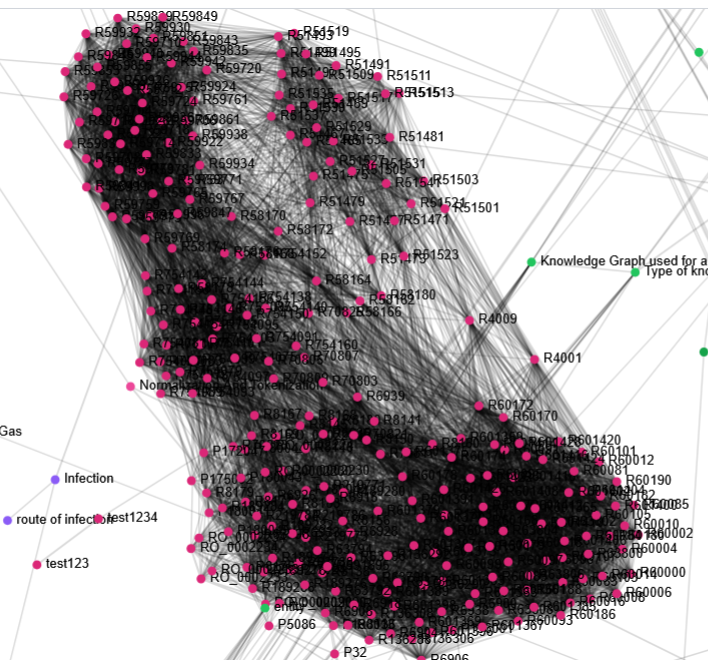}}
    \caption{Cluster containing IDs instead of labels (size 312)}
    \label{fig:ID_cluster}
\end{minipage}
\hspace{0.3cm}
\begin{minipage}{0.5\textwidth}
    \centering
    \fbox{\includegraphics[width=\linewidth, trim=0 0.52cm 0 0.52cm,
    clip]{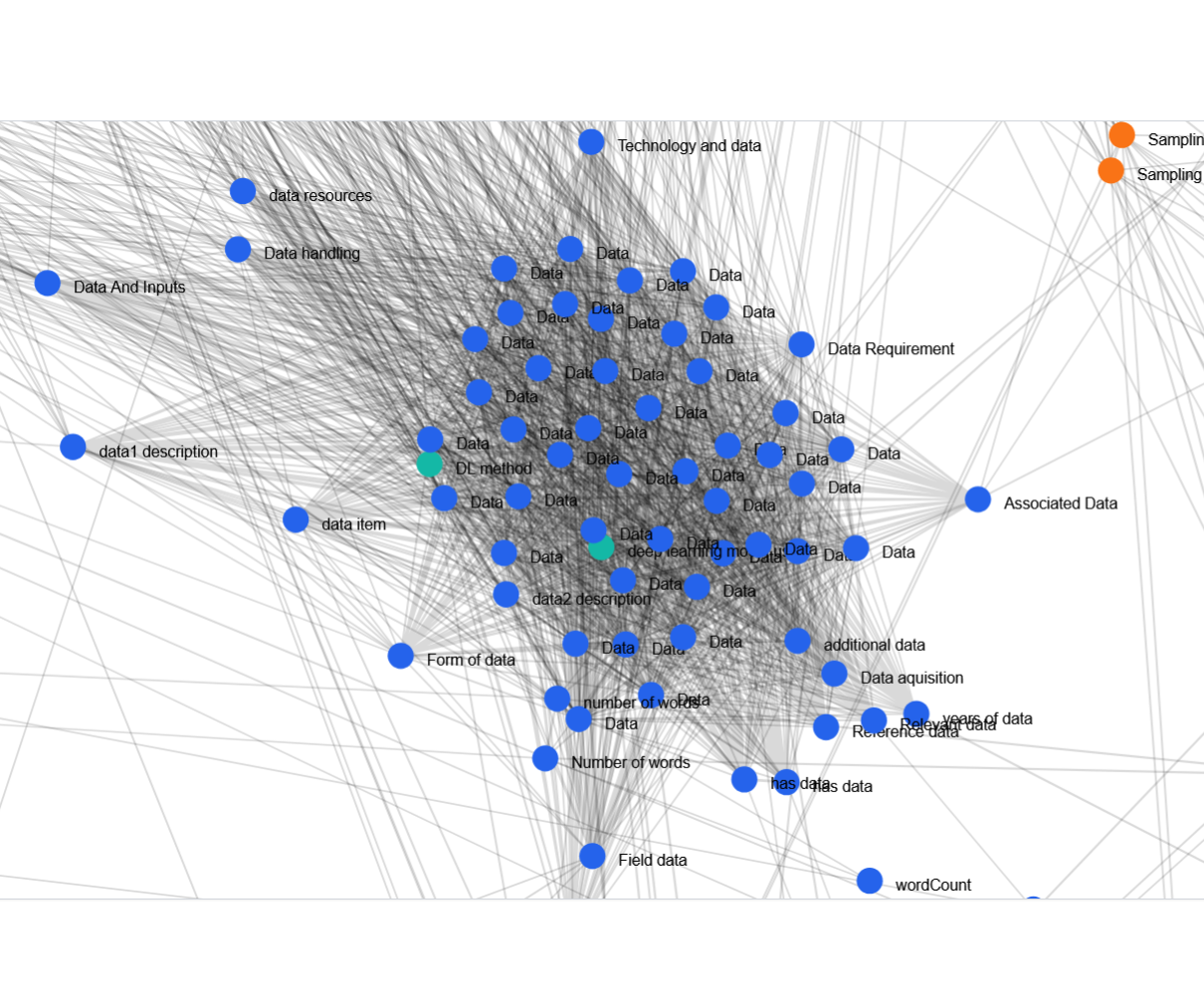}}
    \caption{Example cluster describing data predicates (size 122)}
    \label{fig:semantic_cluster}
\end{minipage}
\vspace{-0.5cm}
\end{figure}

\noindent\textit{Use of identifiers instead of labels.} 
In some cases, users (accidentally) refer to predicates via IDs rather than their labels (see \autoref{fig:ID_cluster}). This introduces artificial lexical similarity, causing unrelated predicates to be grouped into a large cluster. These are false positives caused by label-based similarity, highlighting a core problem in the RKG interface, making prevention mechanisms urgent.

\noindent\textit{Lack of standardization for frequently used predicates.} We also observe clusters containing semantically similar predicates (see \autoref{fig:semantic_cluster}) created by different users for common modeling concepts (e.g., \textit{method}, \textit{data}, \textit{material}). These are not always true duplicates. Resolving them requires human judgment and contextual inspection of predicate usage.

\vspace{-0.5cm}
\begin{mdframed}
     \hypertarget{takeaway}{\textbf{Takeaway:}} The analysis reveals $30\%$ predicate redundancy in the ORKG, indicating significant resolution potential. Cluster sizes are highly skewed, with many singletons and few large clusters. Recurring modeling patterns expose sources of redundancy and require different prevention mechanisms.
\end{mdframed}

\section{Discussion}\label{ch:Discussion}
This section will reflect on the implementation of the framework, present limitations and implications.
The proposed framework is implemented in the ORKG context, with DD, DR, and prototypical DP functionalities integrated into \href{https://scikgdash.orkg.org/predicates}{\textcolor{link}{SciKGDash}}. Supported resolution actions include soft-merging, hard-merging, and deletion, while distinction is implicitly handled via ORKG redirects.

\noindent\textbf{Reflection.} The similarity graph view enables exploration of predicate clusters. Inspecting individual clusters in detail includes usage context and automatically generated suggestions for resolution actions. We identified three recurring modeling practices (cf. \autoref{ch:Evaluation}) that contribute to duplicate predicates: Repeated creation of identical predicates by the same user, the use of identifiers instead of labels, and the lack of standardization for frequently used predicates.
These observations provide concrete starting points for prevention mechanisms. The confusion between IDs and labels is partly rooted in the ORKG interface, where searching by ID requires a preceding \#-symbol, a non-transparent feature that increases the likelihood of unintended duplicate creation. Likewise, the repeated creation of identical predicates by the same user points to missing feedback during predicate creation~\cite{DSouza.2024}. Currently, predicates can be created even if identical labels already exist, indicating the absence of basic guardrails such as duplicate warnings or recommendations.
More restrictive approaches for controlled vocabularies, such as limiting predicate creation to curators or introducing a formal submission workflow (as in \href{https://www.wikidata.org/wiki/Wikidata:Property_creation}{\textcolor{link}{Wikidata}}), could mitigate these issues but must be balanced against ORKG’s openness as a crowdsourced platform~\cite{Karras.2021,Nechakhin.2024}. 
Predicates can also be introduced through multiple entry points (e.g., frontend, REST API, Python package), complicating DP since consistent validation must be enforced across all pathways.
In this context, mapping strategies become essential, particularly when importing data from external tools~\cite{John_ExtracTable.2025,Nechakhin.2024}. Approaches from ontology alignment can be leveraged to match incoming predicates to existing ones. Prior work on semi-automated predicate mapping for CSV import~\cite{John_Scimantify.2025} can be extended using embedding-based similarity as part of a preventive pipeline (cf. \autoref{ch:Implementation}).
The evaluation suggests that up to $30\%$ of predicates may be redundant and potentially resolvable. The aim is to reach 0\% redundancy. Although based on automated suggestions requiring human validation, they highlight substantial potential for improving RKG quality. To motivate continuous cleanup, \href{https://scikgdash.orkg.org/predicates}{\textcolor{link}{SciKGDash}} computes a lightweight coherence score based on cluster sizes that increases as clusters are reduced through deletion and merge actions. The aim is to reach 100\% coherence.
At the same time, the cluster size distribution shows that the majority of clusters are singletons. These reflect either semantically distinct predicates or similarities below the threshold ($0.65$). This is further reinforced by the \href{https://scikgdash.orkg.org/description-coverage}{\textcolor{link}{prior observation in SciKGDash}} that $76.62\%$ of predicates lack descriptions, which reduces the effectiveness of embedding-based similarity and may lead to missing connections between semantically related predicates.\vspace{0.2cm} 

\noindent\textbf{Limitations.}
Several limitations affect both the framework and the evaluation. \textit{First}, the quality of DD depends on the chosen embedding and clustering strategy, making both, the generated candidate groups and the reported redundancy estimates sensitive to missing or noisy textual metadata, similarity thresholds, and parameter choices.
\textit{Second}, clustering is performed as a preprocessing step, constraining subsequent DR actions and the evaluation to predicates within the same cluster. Consequently, semantically related predicates assigned to different clusters cannot be merged or counted as duplicate predicate candidates, potentially underestimating redundancy in ORKG. This reflects a design trade-off favoring high recall during clustering, followed by refinement within clusters. 
\textit{Third}, the proposed resolution actions rely on heuristic rules and do not incorporate domain-specific semantics or contextual reasoning. Consequently, automated suggestions may include false positives, affecting the reliability of the reported redundancy estimates, and require human validation. 
\textit{Finally}, the evaluation itself is limited to a snapshot of the ORKG and does not assess the long-term impact of applying DR strategies or the effectiveness of DP mechanisms in practice. In addition, the framework is shaped by ORKG-specific modeling patterns, limiting its generalizability to other RKGs.

\noindent\textbf{Implications.}
Despite these limitations, the results have several important implications. \textit{First}, predicate deduplication can significantly improve RKG quality. Consolidating predicates leads to more consistent relation usage, which facilitates entity alignment and entity DD, reduces semantic fragmentation, and improves downstream applications such as querying and analysis. \textit{Second}, these findings highlight curation as a core component of RKG design. Because predicates encode semantics rather than real-world entities, their consolidation cannot be fully automated and requires human judgment. This positions curators as key users and underscores the need for dedicated interfaces, such as curation dashboards, to support inspection, interpretation, and decision-making. The proposed framework (cf. \autoref{ch:Framework}) enables the integration of deduplication workflows into RKG infrastructures and elevates predicate redundancy to a core data quality challenge rather than a downstream issue.
\textit{Third}, duplicate creation emerges not only as a technical issue, but as a consequence of user interaction and interface design. The absence of validation during predicate creation represents a critical gap in ORKG’s crowdsourced modeling approach. Prevention mechanisms should combine backend validation with user-facing guidance, including improved autocomplete, similarity-based recommendations during creation, and clearer modeling guidelines. Integrating DD directly into the creation workflow would allow redundancy to be addressed at its source, reducing the need for costly post-hoc cleanup. In this sense, DD becomes an integral component of DP within the iterative framework (cf. \autoref{fig:framework}). \textit{Finally}, the combination of DD, DR, and DP suggests a shift from reactive to proactive RKG quality management. 

\section{Conclusion}\label{ch:Conclusion}
This section will summarize the paper and outlines future work opportunities.

\noindent\textbf{Summary.} We presented a framework for duplicate detection (DD), resolution (DR), and prevention (DP), demonstrated for the Open Research Knowledge Graph (ORKG). By combining embedding-based similarity, clustering, and heuristic-based action suggestions, the framework identifies and consolidates duplicate predicates. Its integration into the deployed curation dashboard \href{https://scikgdash.orkg.org/predicates}{\textcolor{link}{SciKGDash}} shows how these methods can be operationalized through interactive interfaces that support curators in inspecting and resolving data quality challenges.

Our evaluation shows that up to $30\%$ of predicates may be redundant, highlighting substantial potential for improving Research Knowledge Graph (RKG) quality and consistency. At the same time, the predominance of singleton clusters suggests that many predicates are already well-formed or fall below current similarity thresholds. Qualitative analysis revealed recurring modeling patterns, including repeated creation of identical predicates by the same user, confusing identifiers with labels, and lack of standardization
for frequently used predicates.
These findings show that duplicate predicates arise from both user behavior and system design. Addressing them requires a holistic approach combining automated detection, human-centered curation, and preventive mechanisms. The proposed framework links these aspects into an iterative data quality management process for RKGs.

\noindent\textbf{Future Work.}
Several directions can extend this work. \textit{First}, improving scalability of DD is essential~\cite{Huaman.2020}, e.g., by replacing pairwise similarity with approximate nearest neighbor search~\cite{Chen.2023}. Future work should also benchmark runtime, analyze pipeline steps efficiency, and evaluate the Leiden algorithm~\cite{Traag.2019} as an alternative to Louvain for improved efficiency and clustering quality.
\textit{Second}, incorporating richer semantic signals, such as domain and range analysis of predicates, may improve both DD and DP. While ORKG does not enforce such constraints, usage patterns may implicitly reveal them. 
User studies could evaluate whether these signals improve modeling decisions.
Analyzing co-occurring predicates could suggest existing predicates for a given domain or modeling scope~\cite{Nechakhin.2024}.
\textit{Third}, prevention mechanisms should be more tightly integrated into the predicate creation process. Completing the similarity-based autocomplete in the frontend can guide users toward existing predicates and reduce duplicate creation.
\textit{Fourth}, large language models (LLMs) offer opportunities to support curation tasks. They can support the distinguish action by suggesting labels, generating descriptions from usage context, explaining predicate differences for human decision-making, and automating trivial duplicate cases based on curation history.
\textit{Fifth}, supporting cross-cluster actions and more flexible merging strategies would improve the DR process. Current cluster-level constraints may prevent consolidating semantically related predicates, that were not grouped initially.
\textit{Finally}, integrating the framework into real-world curation workflows remains a key step. Collaborating with the ORKG Curation \& Community Building team will enable a human gold standard for practical validation, including clustering accuracy, precision/recall and parameter sensitivity analyses, and clustering parameters tuning.

Overall, this work lays the foundation for treating duplicate predicate management as a core aspect of RKG curation, moving from reactive cleanup toward proactive and continuous RKG quality management.

%
%
%
\bibliographystyle{splncs04}
\bibliography{references}

\end{document}